\documentclass[a4paper,11pt]{article}
\pdfoutput=1 

\usepackage{jheppub} 

\usepackage[T1]{fontenc} 

\title{\boldmath Parafermionization, bosonization, and critical parafermionic theories}

\author[a,b,1]{Yuan Yao,\note{Corresponding author.}}
\author[b,c]{Akira Furusaki}

\affiliation[a]{Institute for Solid State Physics, University of Tokyo, \\Kashiwa, Chiba 277-8581, Japan}
\affiliation[b]{Condensed Matter Theory Laboratory, RIKEN CPR, \\Wako, Saitama 351-0198, Japan}
\affiliation[c]{Quantum Matter Theory Research Team, RIKEN CEMS, \\Wako, Saitama 351-0198, Japan}

\emailAdd{yuan.yao@riken.jp}

\abstract{We formulate a $\mathbb{Z}_k$-parafermionization/bosonization scheme for one-dimensional lattice models and field theories on a torus, 
starting from a generalized Jordan-Wigner transformation on a lattice, which extends the Majorana-Ising duality at $k=2$. 
The $\mathbb{Z}_k$-parafermionization enables us to investigate the critical theories of parafermionic chains whose fundamental degrees of freedom are parafermionic, and we find that their criticality cannot be described by any existing conformal field theory. 
The modular transformations of these parafermionic low-energy critical theories as general consistency conditions are found to be unconventional in that their partition functions on a torus transform differently from any conformal field theory when $k>2$.
Explicit forms of partition functions are obtained by the developed parafermionization for a large class of critical $\mathbb{Z}_k$-parafermionic chains, whose operator contents are intrinsically distinct from any bosonic or fermionic model in terms of conformal spins and statistics.
We also use the parafermionization to exhaust all the $\mathbb{Z}_k$-parafermionic minimal models, complementing earlier works on fermionic cases.}

\begin{document} 
\maketitle
\flushbottom

\section{Introduction}
As a central task in quantum many-body theory, various bosonic and fermionic phases with/without symmetries have been identified and classified in recent decades, notable examples being topological-phase classifications~\cite{Chen:2010aa,Wen:TOreview2013,Gu:2009aa,Pollmann:2012aa} beyond the Landau-Ginzburg symmetry-breaking paradigm. 
Perhaps the simplest generalizations of bosonic or fermionic degrees of freedom
are parafermions, which are defined by generalized fractional commutation relations and emerge on the edges of two-dimensional fractional quantum phases~\cite{Lindner:2012aa,Cheng:2012aa,Vaezi:2013aa,Barkeshli:2013aa,Barkeshli:2013ab,Mong:2014aa,Khan:2014aa,Alicea:2016aa,Santos:2017aa}.
Many efforts have been also done on the classification of gapped parafermionic topological orders without symmetry~\cite{Fendley:2012aa,Bondesan:2013aa,Motruk:2013aa,Alicea:2016aa,Alexandradinata:2016aa,Iemini:2017aa} and with symmetries~\cite{Meidan:2017aa}. 

Parafermions naturally emerge in critical $\mathbb{Z}_k$-clock models~\cite{Fradkin:1980aa,Zamolodchikov:1985aa,Gepner:1987aa,Fendley:2012aa} which are made of generalized bosonic spins. 
The typical example is the well-known critical quantum transverse Ising model ($k=2$), which is ``equivalent'' to a massless Majorana fermionic chain, i.e., a $\mathbb{Z}_2$ parafermion~\cite{Francesco:2012aa}. 
However, 
the Majorana fermions and the Ising spins are significantly different in nature in that the Majorana fermions, which are local excitations in Majorana system and obey fermion statistics, are forbidden to exist in the local excitation content of the Ising chain, whose local excitations are bosonic.
Thus, more precisely speaking, 
the massless Majorana fermion chain is actually equivalent to a proper stacking~\cite{Kapustin:2017aa,Karch:2019aa,Yao:2019ab,Hsieh:2020aa} of the critical Ising chain and a gapped Kitaev chain in its $\mathbb{Z}_2$-topologically nontrivial phase~\cite{Kitaev:2001aa} providing the fermionic nature.
Therefore,
the critical theories of fermions (e.g., the critical theory of Majorana chains) 
are called ``fermionic conformal field theories'' to be distinguished from the critical bosonic theories, i.e., the so-called conformal field theories (CFTs).
The classification~\cite{Ingo:2020aa} and the minimal and rational models of fermionic CFTs~\cite{Hsieh:2020aa,Kulp:2020aa,Gaiotto:2020aa,Bae:2020aa} are of intense interest recently. 

In fact,
the construction of the general critical theory of the $\mathbb{Z}_{k}$-parafermionic systems still remains an open problem for $k>2$,
and the significant distinctions of parafermion statistics from boson and fermion statistics imply that these parafermionic critical theories with $k>2$ may not be described by any existing CFT, e.g., bosonic or fermionic CFTs.
In this respect we note that
there have been progress on the concept called ``parafermionic CFTs''~\cite{Zamolodchikov:1985aa,Gepner:1987aa,Li:2015aa,Lahtinen:2017aa}, which are field theories containing parafermionic operators, but still obeying the conventional modular invariance of bosonic CFTs.
Indeed, 
they are found to describe the bosonic systems like $\mathbb{Z}_k$ clock models~\cite{Gepner:1987aa,Li:2015aa,Lahtinen:2017aa}, rather than genuinely parafermionic systems of our current interest whose fundamental degrees of freedom are parafermions.
Therefore, the study on the (genuinely) parafermionic critical systems and the fundamental constraints on their low-energy effective theories is important and interesting in its own right and enriches the general framework and methodology of CFTs. 
Furthermore,
the classification and solution of minimal and rational~\cite{Cappelli:1987aa,Kato:1987aa,Francesco:2012aa} $\mathbb{Z}_{k}$-parafermionic models beyond $k=2$ were still lacking due to the absence of full understanding on parafermionic systems.

In this work, 
we investigate the fundamental properties of general critical theories of underlying parafermionic chains and their relation to parafermionic topological phases.
As a main method of our study, 
we first develop a (1+1)-dimensional parafermionization together with a bosonization as its inverse to relate a parafermionic theory to a bosonic theory by a one-to-one correspondence. 
It can be also regarded as an attachment construction using a nontrivial topological phase of a parafermionic chain~\cite{Fendley:2012aa,Bondesan:2013aa,Motruk:2013aa,Alicea:2016aa,Alexandradinata:2016aa,Iemini:2017aa} generalizing the Kitaev-chain attachment argument in $k=2$~\cite{Karch:2019aa,Yao:2019ab,Hsieh:2020aa}.
From this viewpoint,
parafermionic chains and bosonic chains are expected to be indistinguishable locally since their differences result from this global topological factor.
The parafermionization method also enables us to study the general properties of partition functions of critical parafermionic chains, which obey unconventional modular transformations, 
distinct from any existing bosonic or fermionic CFT. 
The source of this unconventional modular invariance is also interpreted from a lattice viewpoint, and we propose it as a general consistency condition on any critical theory of parafermionic systems. 
We also apply the parafermionization to explicitly calculate the partition function of a large class of critical parafermionic chains, 
from which their intrinsic fractional statistics can be read off.

The paper is organized as follows. 
In Sec.~\ref{f-k trans}, we first introduce a generalized Jordan-Wigner transformation to obtain the fermionization. 
Next, it is re-interpreted as the attachment construction in Sec.~\ref{attach}. 
The unconventional modular transformation is investigated in Sec.~\ref{convention}. 
Then, 
we show that the CFTs obeying conventional modular invariances cannot correctly describe the lattice model of critical parafermionic phases in Sec.~\ref{invertible}.
Finally, we exhaust the remaining $\mathbb{Z}_{k>2}$-parafermionic minimal models beyond solved bosonic/fermionic cases,
and discuss a large class of {minimal/nonminimal parafermionic systems to quantify} their parafermionic statistics in Sec.~\ref{fractional_stat}, 
and conclude in Sec.~\ref{conclusion}.
Appendices include CFTs on a torus reviewed in Appendix~\ref{lattice_torus},
a detailed discussion on the state-operator correspondence in Appendix~\ref{locality}, and partition functions of the $\mathbb{Z}_{k>2}$-parafermionic minimal models in Appendix~\ref{calculation}.

\section{Fradkin-Kadanoff transformation and boundary conditions}
\label{f-k trans}
Let us consider a quantum $\mathbb{Z}_k$-generalization of Ising degrees of freedom or $\mathbb{Z}_k$-spin, $\sigma_j$ and $\tau_j$, at each site $j$ in a one-dimension lattice:
\begin{eqnarray}
\label{k_spin}
\sigma_j^k=\tau_j^k=1,\,\,\sigma^\dagger_j\equiv\sigma_j^{-1},\,\,\tau^\dagger_j\equiv\tau_j^{-1},\,\,\tau_j\sigma_j\tau_j^{-1}=\omega^*\sigma_j, 
\end{eqnarray}
where $\omega\equiv\exp(i2\pi/k)$, and
$\sigma$'s and $\tau$'s are bosonic in the sense that they all commute at different sites and the local Hilbert space at the site $j$ is (minimally) $k$-dimensional.
They are local operators that can be represented as
\begin{equation}
\sigma_j=\cdots\otimes 1_k \otimes
\left(\begin{array}{ccccc}
0 & 1 & 0 & \ldots & 0\\
0 & 0 & 1 & \ldots & 0\\
\vdots & \vdots & \vdots & \ddots & \vdots\\
0 & 0 & 0 & \ldots & 1\\
1 & 0 & 0 & \ldots & 0
\end{array}\right)
\otimes 1_k \otimes\cdots,\,\,
\tau_j=\cdots\otimes 1_k \otimes
\left(\begin{array}{ccccc}
1 & 0 & 0 & \ldots & 0\\
0 & \omega & 0& \ldots & 0\\
\vdots & \vdots & \vdots & \ddots & \vdots\\
0 & 0 & 0 & \ldots & \omega^{k-1}
\end{array}\right)
\otimes 1_k \otimes\cdots ,
\end{equation}
where $1_k$ is the $k\times k$ unit matrix and the matrices other than $1_k$ appear only in the $k$-dimensional local Hilbert space at the site $j$.

Such a generalized $\mathbb{Z}_k$-spin picture in a finite chain is, \textit{roughly speaking}, equivalent to a parafermionic system by the following Fradkin-Kadanoff transformation~\cite{Fradkin:1980aa} generalizing the Jordan-Wigner transformation: 
\begin{eqnarray}
\gamma_{2j-1}\equiv\sigma_j\prod_{i<j}\tau_i;\,\,\,\,\gamma_{2j}\equiv\omega^{(k-1)/2}\sigma_j\prod_{i\leq j}\tau_i, 
\end{eqnarray}
where, for a finite chain $j=1,2,\cdots,L$, the product terminates at the most ``left'' site $1$:
\begin{eqnarray}
\label{f_k}
\gamma_{2j-1}\equiv\sigma_j\prod_{i=1}^{j-1}\tau_i;\,\,\,\,\gamma_{2j}\equiv\omega^{(k-1)/2}\sigma_j\prod_{i=1}^j\tau_i .
\end{eqnarray} 
{Note that the product for $\gamma_{2j}$ includes $\tau_j$.}
As a generalization of Majorana fermions ($k=2$), the parafermionic degrees of freedom satisfy
\begin{eqnarray}\label{para_1}
\gamma_j^{k}=1;\,\,\gamma^\dagger_j=\gamma^{-1};\,\,\gamma_j\gamma_l=\omega^\text{sgn$(l-j)$}\gamma_l\gamma_j\,\,\,(l\neq j), 
\end{eqnarray}
which we will take as a defining feature of the parafermionic chains and we can forget about the bosonic model $\{\sigma_j,\tau_j\}$ we started with. 
The Hilbert space of the parafermionic chain is defined to be the same as that of the following auxiliary local $\mathbb{Z}_k$-spins
\begin{eqnarray}\label{inverse_f_k}
\tilde{\sigma}_j\equiv\gamma_{2j-1}\prod_{i<j}\left(\omega^{(k-1)/2}\gamma_{2i}^\dagger\gamma_{2i-1}\right);\,\,\tilde{\tau}_j\equiv\omega^{(1-k)/2}\gamma_{2j-1}^\dagger\gamma_{2j},
\end{eqnarray}
which turns out to be the inverse of the Fradkin-Kadanoff transformation~(\ref{f_k}).
However, 
the parafermionic and the dual bosonic models are intrinsically different in the definition of locality of operators~\footnote{Here the intrinsic non-locality of fermions is reflected by the fact that the (para)fermion operators cannot be represented by any local matrix (without string operators) no matter how we choose the basis. Therefore, $\mathbb{Z}_k$ is always a global symmetry of general parafermionic Hamiltonians.}, 
while they still share the same Hilbert space.
When $k>2$, there is one additional significant aspect in a finite chain with sites $j=1,2,\cdots,L$ as follows. 
The last relation in Eq.~(\ref{para_1}) signifies the absence of a unitary translation symmetry $U$ such that $U\gamma_{j}U^\dagger=\gamma_{j+1}$ if $j=1,2,\cdots,2L-1$ and $U\gamma_{2L}U^\dagger=\gamma_1$,
because $U$, if existing, acting on both sides of $\gamma_{2L}\gamma_1=\omega^*\gamma_1\gamma_{2L}$ would be inconsistent with $\gamma_{1}\gamma_2=\omega\gamma_2\gamma_1$ when $\omega^*\neq\omega$, i.e., $k>2$.
Therefore,
the branch cut connecting $\gamma_{2L}$ and $\gamma_1$ is not equivalently oriented with the other links between $\gamma_j$ and $\gamma_{j+1}$ when $k>2$.
Nevertheless, 
we can still define a different translation transformation $V_\text{transl}$ by the help of the auxiliary bosonic spins conversely defined by the parafermions as in Eq.~(\ref{inverse_f_k}), satisfying
\begin{eqnarray}\label{transl_b}
V_\text{transl}\tilde{\sigma}_j V^\dagger_\text{transl}=\tilde{\sigma}_{j+1};\,\,V_\text{transl}\tilde{\tau}_j V^\dagger_\text{transl}=\tilde{\tau}_{j+1}
\end{eqnarray}
with $(\tilde{\sigma}_{L+1},\tilde{\tau}_{L+1})\equiv(\tilde{\sigma}_1,\tilde{\tau}_1)$, which completely determines, by the relation~(\ref{f_k}), the action of $V_\text{transl}$ on the original parafermionic chain $\{\gamma_j\}$ with a finite length $2L$ as:
\begin{equation}
\begin{split}
\label{transl}
V_\text{transl}\gamma_{2j-1} V^\dagger_\text{transl}&=\gamma_{2j+1}\left(\omega^{(k-1)/2}\gamma_2^\dagger\gamma_1\right),\\
V_\text{transl}\gamma_{2j} V^\dagger_\text{transl}&=\gamma_{2j+2}\left(\omega^{(k-1)/2}\gamma_2^\dagger\gamma_1\right), 
\end{split}
\end{equation}
for the parafermions at $j=1,2,\cdots,L-1$,
while, for an infinitely long chain $j=\cdots,-2,-1,0,1,\cdots$, the finite-length corrections ($\omega^{(k-1)/2}\gamma_2^\dagger\gamma_1 =\tilde{\tau}_1^\dagger)$ coming from the leftmost site disappear.
Additionally,
the boundary parafermions transform as,
\begin{equation}
\begin{split}
V_\text{transl}\gamma_{2L-1} V^\dagger_\text{transl}&=\gamma_{1}\left(\omega^{(k-1)/2}\gamma_2^\dagger\gamma_1\right)Q_f,\\
V_\text{transl}\gamma_{2L} V^\dagger_\text{transl}&=\gamma_{2}\left(\omega^{(k-1)/2}\gamma_2^\dagger\gamma_1\right)Q_f,
\end{split}
\end{equation}
{where $Q_f$ is a generator of global $\mathbb{Z}_k$ symmetry defined below.}
Such a translation transformation also keeps the algebra~(\ref{para_1})
since, if we define ${\gamma}'_j=V_\text{transl}\gamma_jV_\text{transl}^\dagger$, then ${\gamma}'_j$'s satisfy the same algebra~(\ref{para_1}) by replacing $\gamma_j\mapsto{\gamma}'_j$. 
Thus, $V_\text{transl}$ is the appropriate translation transformation for parafermionic chains of a finite or infinite length,
and it will be used when we formulate the parafermions on a general space-time torus later in Sec.~\ref{convention} and Appendix~\ref{lattice_torus}.

We will use the subscript ``$b$'' to label the bosonic spin system (\ref{k_spin}) and ``$f$'' for the parafermionic system~(\ref{para_1}).
In either picture, there exist global $\mathbb{Z}_k$ symmetries: $Q_f\gamma Q^{-1}_f=\omega^*\gamma$ and $Q_b\sigma Q_b^{-1}=\omega^*\sigma$ generated by: 
\begin{eqnarray}
\label{compare}
Q_f&\equiv&\prod_{i}\left[\omega^{(1-k)/2}\gamma_{2i-1}^\dagger\gamma_{2i}\right];\,\,\,Q_b\equiv\prod_i{\tau}_i, 
\end{eqnarray}
and they are the same $Q_f=Q_b$ by Eq.~(\ref{f_k}).
In this paper, we will focus only on parafermionic systems with such $\mathbb{Z}_k$ symmetries. 

Now we derive the exact correspondence between the parafermionic chains and $\mathbb{Z}_k$-spin bosonic chains of finite length $L$ under twisted boundary conditions.
Without loss of generality (see the discussion later), we consider the nearest-neighbor coupling and compare the edge-closing term 
\begin{eqnarray}
\label{compare_1}
\sigma^\dagger_1\sigma_L\omega^{- a_1 }&=&\left[\omega^{-(k+1)/2}\gamma_1^\dagger\gamma_{2L}\right](\omega^{ a_1 -1} Q_b)^\dagger\nonumber\\
&=&\left[\omega^{-(k+1)/2}\gamma_1^\dagger\gamma_{2L}\right](\omega^{q_b+ a_1 -1} )^*,
\end{eqnarray}
with the corresponding terms in the bulk: 
\begin{eqnarray}\label{nearest}
\sigma_{j+1}^\dagger\sigma_j&=&\left[\omega^{-(k+1)/2}\gamma_{2j+1}^\dagger\gamma_{2j}\right],\,\,~(1\leq j\leq L-1),
\end{eqnarray}
where $ a_1 $ is a mod-$k$ integer-valued parameter specifying a $\mathbb{Z}_N$-twisted boundary condition of the $\mathbb{Z}_k$-spin chain imposed by
\begin{equation}
\sigma_{L+1}=\omega^{a_1}\sigma_1,
\label{twisted bc for boson}
\end{equation}
and we have restricted to the bosonic Hilbert subspace by the $\mathbb{Z}_k$ symmetry: $Q_b=\omega^{q_b}$ where $q_b$ is defined modulo $k$. 
From Eq.~(\ref{compare_1}) and $Q_f=Q_b$, 
we obtain the following mapping between the Hamiltonians: 
\begin{eqnarray}\label{correspondence}
\mathcal{H}_f(s_1)|_{Q_f=\omega^{q_b}}=H_b(1+s_1-q_b)|_{Q_b=\omega^{q_b}},
\end{eqnarray}
where $\mathcal{H}_f( s_1 )$ denotes the Hamiltonian of the parafermionic chain twisted by $Q_f^{s_1}$ and $H_b(a_1)$ for the bosonic chain twisted by $Q_b^{a_1}$.
For general edge-closing terms $H_\text{edge}$ with a finite range $l_\text{edge}\ll L$, we can define the unit twisting as acting $Q_\text{edge}\equiv\prod_{i=1}^{l_\text{edge}}\tau_i=\prod_{i=1}^{l_\text{edge}}\left[\omega^{(1-k)/2}\gamma_{2i-1}^\dagger\gamma_{2i}\right]$ on $H_\text{edge}$ as $Q_\text{edge}^{-1}H_\text{edge}Q_\text{edge}$.
From Eq.~(\ref{nearest}),
we expect that the parafermionic chain and its bosonic dual obtained by Eq.~(\ref{inverse_f_k}) are locally indistinguishable since, on an infinitely long chain $j=\cdots,-1,0,1\cdots$ without boundaries, 
the Hilbert-space dependent boundary twistings are irrelevant.
We will see that the global aspect of the difference can be understood by a topological-phase attachment in Sec.~\ref{attach}.

The partition-function correspondence with the inverse temperature $\beta$ as the imaginary time can be obtained as
\begin{eqnarray}
\label{para}
\mathcal{Z}_{s_1,s_2}&\equiv&\text{Tr}\{(Q_f)^{1+ s_2 }\exp[-\beta \mathcal{H}_f( s_1 )]\}\nonumber\\
&=&\sum_{q_b=0}^{k-1}\text{Tr}|^{}_{Q_f=\omega^{q_b}}\{(Q_f)^{1+ s_2 }\exp[-\beta \mathcal{H}_b( 1+s_1-q_b )]\}\nonumber\\
&=&\sum_{q_b=0}^{k-1}\text{Tr}\{\mathcal{P}_{q_b}\,(Q_b)^{1+ s_2 }\exp[-\beta \mathcal{H}_b( 1+s_1-q_b )]\}\nonumber\\
&=&\frac{1}{k}\sum_{q_b=0}^{k-1}\text{Tr}\left\{\sum_{p=0}^{k-1}\left(\omega^{-q_b}Q_b\right)^p(Q_b)^{1+ s_2 }\exp[-\beta \mathcal{H}_b( 1+s_1-q_b )]\right\}\nonumber\\
&=&\frac{1}{k}\sum_{ a_1 , a_2 }\omega^{(1+ s_1 - a_1 )(1+ s_2 - a_2 )}\text{Tr}\left\{(Q_b)^{ a_2 }\exp[-\beta H_b( a_1 )]\right\}\nonumber\\
&=&\frac{1}{k}\sum_{ a_1 , a_2 }\omega^{(1+ s_1 - a_1 )(1+ s_2 - a_2 )}Z_{a_1,a_2},
\end{eqnarray}
by the Hamiltonian correspondence~(\ref{correspondence}) {with $a_{1,2}$ summed from $0$ to $k-1$}, the projection operator onto the Hilbert subspace with $Q_f=\omega^{q_b}$
\begin{eqnarray}
\label{projection_op}
\mathcal{P}_{q_b}\equiv\frac{1}{k}\sum_{p=0}^{k-1}\left(\omega^{-q_b}Q_f\right)^p, 
\end{eqnarray}
the correspondence $Q_f=Q_b$, and
\begin{eqnarray}
Z_{a_1,a_2}\equiv\text{Tr}\{(Q_b)^{ a_2 }\exp[-\beta H_b( a_1 )]\},
\end{eqnarray}
where $\mathcal{Z}_{s_1,s_2}$ and $Z_{a_1,a_2}$ are parafermionic and bosonic partition function under corresponding boundary-condition twistings, and we have inserted $Q_f^{( 1+s_2 )}$ to twist the temporal direction as well. 
Here, the convention of ``$(1+s_2)$'' is made so that $( s_1 , s_2 )$ reduces to the conventional $\mathbb{Z}_2$ spin structure when $k=2$, and we will call it a ``paraspin'' structure.
Additionally,
this convention is convenient in that $s_1$ and $s_2$ are on an equal footing in Eq.~(\ref{para}). 
In the following discussion, 
$a_{1,2}$ and $s_{1,2}$ are all defined mod $k$ and we will keep using curly $\mathcal{Z}$ to denote the parafermionic partition functions and $Z$ for the bosonic partition functions.

\section{Attachment constructions and bosonizations as inverse}
\label{attach}
To manifest the physical meaning of the parafermionization (\ref{para}), we rewrite it as
\begin{eqnarray}\label{parafermionization}
\mathcal{Z}_{s_1,s_2}=\frac{1}{k}\sum_{a_1,a_2}z^{s_1,s_2}_{-a_1,-a_2}\,Z_{a_1,a_2},
\end{eqnarray}
where the coefficient is defined as
\begin{eqnarray}
\label{kitaev}
z^{s_1,s_2}_{a_1,a_2}\equiv\exp\left[\frac{2\pi i}{k}(1+s_1+a_1)(1+s_2+a_2)\right].
\end{eqnarray}
When $k=2$, $z^{s_1,s_2}_{a_1,a_2}$ reduces to the partition function of the nontrivial topological phase of the Kitaev chain as the $\mathbb{Z}_2$-Arf invariant~\cite{Karch:2019aa,Yao:2019ab,Hsieh:2020aa}, coupled with a background $\mathbb{Z}_2$-gauge field $(a_1,a_2)$.
{Additionally,
when both the parafermionic and the bosonic theories are coupled to dynamic $\mathbb{Z}_k$-gauge fields,
the resultant gauged theories are the same modular invariant theory with the orbifold partition function $Z_\text{orb}=k^{-1}\sum_{s_{1,2}}\mathcal{Z}_{s_1,s_2}=k^{-1}\sum_{a_{1,2}}{Z}_{a_1,a_2}$ generalizing $k=2$ cases~\cite{Yao:2019ab}.}

\subsection{Attachment of a gapped parafermionic chain}
Here, we will argue that Eq.~(\ref{kitaev}) is exactly the partition function of a generalized Kitaev phase by using the results from Ref.~\cite{Alexandradinata:2016aa}, where it is shown that the open parafermionic chain
\begin{eqnarray}
\mathcal{H}_\text{open}=-\frac{1}{k}\sum_{j=1}^{L-1}\sum_{p=0}^{k-1}\left\{\left[\omega^{(k-1)/2}\gamma^\dagger_{2j+1}\gamma_{2j}\right]^p-1\right\}
\end{eqnarray}
has $k$-fold degenerate gapped ground states representing the dangling edge modes and having $\mathbb{Z}_k$ charges as $Q_f=\omega^{q_f}$ with $q_f=0,1,\cdots,k-1$, separately.
Here, the additional constant ``$-1$'' in $\mathcal{H}_\text{open}$ is to normalize the ground-state energy density to be zero.
The complicated polynomial summation will be useful for closing the chain as we will see later.
In addition, the lattice model above is exactly solvable since the nearest-neighbor hoppings commute with each other and thus the energy of ground states can be saturated by
\begin{eqnarray}
\left.\omega^{(k-1)/2}\gamma_{2j+1}^\dagger\gamma_{2j}\right|_\text{G.S.}=1, 
\end{eqnarray}
which implies that, by Eq.~(\ref{compare}),
\begin{eqnarray}
Q_f|_\text{G.S.}=\omega^{(1-k)/2}\gamma^\dagger_1\gamma_{2L}.
\end{eqnarray}
On the other hand, 
we can extract the charge sector with $Q_f=\omega^{q_f}$ by the projection operator~(\ref{projection_op}).
Thus, in the ground-state sector, 
we have
\begin{eqnarray}
\mathcal{P}_{q_f}|_\text{G.S.}=\mathcal{P}_{q_f}'\equiv\frac{1}{k}\sum_{p=0}^{k-1}\left[\omega^{(1-k-2q_f)/2}\gamma^\dagger_1\gamma_{2L}\right]^p.
\end{eqnarray}
Since only the $q_f$ sector takes a nonzero positive value of $\mathcal{P}'_{q_f}$, 
we can gap out the other ground state(s) by the following Hamiltonian: 
\begin{eqnarray}
\label{close}
\mathcal{H}_\text{close}&\equiv&\mathcal{H}_\text{open}-\left(\mathcal{P}'_{q_f}-1\right)
\nonumber\\
&=&-\frac{1}{k}\sum_{p=0}^{k-1}\left\{\sum_{j=1}^{L-1}\left\{\left[\omega^{(k-1)/2}\gamma^\dagger_{2j+1}\gamma_{2j}\right]^p-1\right\}+\left[\omega^{(k-1)/2}\gamma^\dagger_{1}\gamma_{2L}\omega^{1-q_f}\right]^p-1\right\}
.
\end{eqnarray}
With the last interedge coupling, 
the model is still exactly solvable and gapped.  
By a direct observation,
Eq.~(\ref{close}) is exactly the $\mathbb{Z}_k$-twisted Hamiltonian by an $s_1=q_f-1$ twisting, i.e., the $\mathbb{Z}_k$ charge of the gapped unique ground state being $q_f=1+s_1$.
Thus, we obtain the partition function as
\begin{eqnarray}\label{tft}
z^{s_1,s_2}&=&\text{Tr}(Q_f)^{1+s_2}\exp[-\beta\mathcal{H}_\text{close}(s_1)]\nonumber\\
&\approx&\exp\left[\frac{2\pi i}{k}(1+s_1)(1+s_2)\right]
\end{eqnarray}
deeply into the gapped phase. 
After coupling it to a background $\mathbb{Z}_k$-gauge field $(a_1,a_2)$, 
the partition function is precisely that in Eq.~(\ref{kitaev}). 

Therefore, we can view the parafermionization (\ref{parafermionization}) as first coupling the bosonic model with the closed parafermionic chain $\mathcal{H}_\text{close}$ in a conjugate $\mathbb{Z}_k$ representation, i.e., having opposite $\mathbb{Z}_k$ charges, and then orbifolding the $\mathbb{Z}_k$ symmetry. 
Such an attachment interpretation generalizes the former $k=2$ cases of bosonization and fermionization by Jordan-Wigner transformations, whereas the sign in the conjugation is irrelevant there and the closed Kitaev chain is exactly solvable deeply in the topological phase.

\subsection{Bosonization as an inverse}
The parafermionization (\ref{parafermionization}) is actually invertible and it is directly proven that
\begin{eqnarray}\label{invertible_1}
\sum_{a_1,a_2}\frac{1}{k}z^{s_1,s_2}_{-a_1,-a_2}\frac{1}{k}\left(z^{s'_1,s'_2}_{-a_1,-a_2}\right)^*=\delta_{s_1,s_1'}\delta_{s_2,s_2'},\nonumber\\
\sum_{s_1,s_2}\frac{1}{k}z^{s_1,s_2}_{-a'_1,-a'_2}\frac{1}{k}\left(z^{s_1,s_2}_{-a_1,-a_2}\right)^*=\delta_{a_1,a_1'}\delta_{a_2,a_2'},
\end{eqnarray}
thereby
\begin{eqnarray}\label{bosonization}
{Z}_{a_1,a_2}=\frac{1}{k}\sum_{s_1,s_2}\left(z^{s_1,s_2}_{-a_1,-a_2}\right)^*\,\mathcal{Z}_{s_1,s_2}
\end{eqnarray}
as a bosonization. 
The significance of the invertible parafermionization and its inverse is emphasized as follows.
Let us suppose that we have a critical parafermionic chain composed by fundamental degrees of freedom as parafermions.
We can use the bosonization~(\ref{bosonization}) to transform it to its bosonic counterpart and study various properties of the bosonic partition function $Z_{a_1,a_2}$
using powerful theoretical methods developed for bosonic systems. 
Finally, we do the parafermionization~(\ref{parafermionization}) to map it back to the original parafermionic picture, thereby obtaining the corresponding properties of the critical parafermionic theories.
As we will see in the next section, such an invertibility of the parafermionization will play an essential role when we investigate the requirement on modular-transformation condition of the effective field theories of critical parafermionic chains (\ref{para_1}).

\section{Modular transformation of critical parafermions}\label{convention}
In this section, 
we will consider general parafermionic chains at criticality, in which no relevant length scale exists except for the divergent correlation length in the thermodynamic limit.
Conformal field theories are powerful tools to describe various universality classes of critical spin models, and their partition functions are modular invariant when the low-energy effective theories are formulated on a space-time torus~\cite{Francesco:2012aa}.
We expect that the field theories of critical parafermionic chains also have a general modular transformation rule on a torus which is parametrized by a complex number,
\begin{eqnarray}\label{tau_torus}
\tau\equiv\tau_1+i\tau_2,
\end{eqnarray}
to be introduced below from the lattice viewpoint.

\begin{figure}
\begin{center}
\includegraphics[width=9cm,pagebox=cropbox,clip]{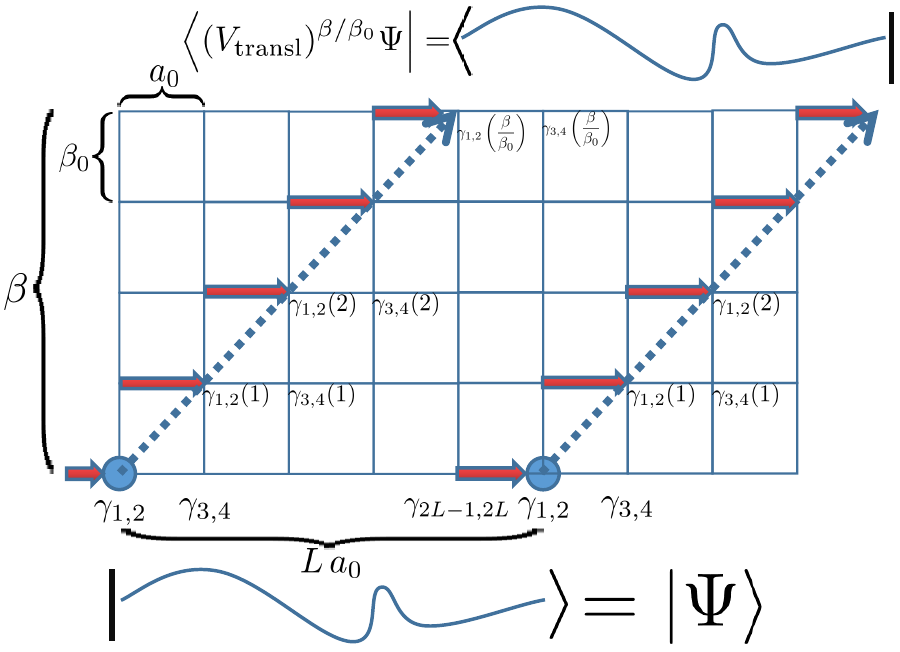}
\caption{The visualization of the partition function $\mathcal{Z}_{s_1,s_2}$: the dash arrow points along the vector $(\tau_1,\tau_2)$ and the state $|\Psi\rangle$ is summed over. 
The solid arrow denotes the $s_1$-twisted link between $\gamma_{2L}$ and $\gamma_1$ which is effectively moved by $V_\text{transl}$ during the time evolution. 
The temporal twisting $Q_f^{s_2+1}$ is not explicitly shown in this figure.
Here we have included $\gamma_{2j-1}$ and $\gamma_{2j}$ into one unit cell thereby denoted by $\gamma_{2j-1,2j}$ for the sake of clarity.}
\label{visual_f}
\end{center}
\end{figure}

To investigate the properties under modular transformation, 
we first define a parafermionic chain on a discrete space-time torus and then take a proper continuum limit.
The procedure is analogous to the bosonic case reviewed in Appendix~\ref{lattice_torus}.
The space-time torus is discretized by introducing a ``lattice'' spacing $\beta_0$ along the imaginary time and the (spatial) lattice spacing $a_0$, which implies that the system length is $La_0$; see FIG.~\ref{visual_f}.
The partition function twisted by $(s_1,s_2)$ in the space-time torus, similarly to Eq.~(\ref{mod_partition}), is defined as
\begin{eqnarray}\label{mod_partition_f}
\mathcal{Z}^\text{latt}_{s_1,s_2}\equiv\text{Tr}\left\{(V^\dagger_\text{transl})^{\beta/\beta_0}[Q_f(\beta)]^{1+s_2}\,\,\mathcal{T}\exp\left[-\int_0^\beta dt\mathcal{H}_f(t;s_1)\right]\right\},
\end{eqnarray}
where $\mathcal{T}$ is the time-ordering operator and $V_\text{transl}$ is defined by Eq.~(\ref{transl}) with the additional ``$1$'' in the exponent $(1+s_2)$ to reproduce the spin-structure convention when $k=2$ as mentioned before.
The imaginary-time evolution is induced by the time-dependent Hamiltonian during the time $t\in[0,(\beta/\beta_0)\beta_0]$: 
\begin{eqnarray}
\mathcal{H}_f(t;s_1)=\left(V_\text{transl}\right)^nH_b(r)\left(V_\text{transl}^\dagger\right)^n,\text{ when }t\in[n\beta_0,(n+1)\beta_0).
\end{eqnarray}
Similarly,
we have also defined a time-dependent $\mathbb{Z}_k$ generator:
\begin{eqnarray}
Q_f(t)=\left(V_\text{transl}\right)^nQ_f\left(V_\text{transl}^\dagger\right)^n,\text{ when }t\in[n\beta_0,(n+1)\beta_0), 
\end{eqnarray}
which is actually time-independent, i.e., $Q_f(t)=Q_f$ by Eq.~(\ref{compare}) and the translation transformation $V_\text{transl}$ in Eq.~(\ref{transl}).
The time evolution~(\ref{mod_partition_f}) is visualized for the discrete space-time in FIG.~\ref{visual_f}.
The additional $(V^\dagger_\text{transl})^{\beta/\beta_0}$ in the partition function~(\ref{mod_partition_f}) effectively moves the state $\langle\Psi|$ involved in the summation by ``Tr'' to match it with the Hamiltonian $\mathcal{H}_f(\beta;s_1)$ as in FIG.~\ref{visual_f}, which is displaced from $\mathcal{H}_f(0;s_1)$ by $a_0\beta/\beta_0$ along the spatial direction.
Thus, the space-time lattice in FIG.~\ref{visual_f} can be seen as a torus with the lateral slope $\beta_0/a_0$.

Then, by the bosonization transformation~(\ref{inverse_f_k}) and its resultant Hamiltonian correspondence~(\ref{correspondence}),
we obtain the correspondence of partition functions
\begin{eqnarray}\label{para_00}
\mathcal{Z}^\text{latt}_{s_1,s_2}&=&\frac{1}{k}\sum_{a_1,a_2}z^{s_1,s_2}_{-a_1,-a_2}\,Z^\text{latt}_{a_1,a_2},
\end{eqnarray}
where $Z^\text{latt}_{a_1,a_2}$ is the partition function of its bosonic dual formulated on the torus in Eq.~(\ref{mod_partition}) rewritten as
\begin{eqnarray}\label{mod_partition_b}
Z^\text{latt}_{a_1,a_2}=\text{Tr}\left\{(V^\dagger_\text{transl})^{\beta/\beta_0}[Q_b(\beta)]^{a_2}\,\,\mathcal{T}\exp\left[-\int_0^\beta dtH_b(t;a_1)\right]\right\}.
\end{eqnarray}
We consider the following standard continuum limit~\cite{Wilson:1974aa,Kogut:1979aa,Sachdev:2007aa,Cardy:2008aa}:
\begin{eqnarray}\label{continuum}
\beta_0,a_0\rightarrow0\text{ with }\beta_0/a_0,\beta\text{ and }La_0\text{ fixed},
\end{eqnarray}
and various lattice couplings are scaled to keep the correlation length $\xi$ fixed, where $\xi$ has the same length dimension as $a_0$ and the (dimensionless) lattice correlation length is $\xi/a_0$.
Then we have the following two dimensionless parameters which are invariant in the continuum limit:
\begin{eqnarray}
\tau_1\equiv\frac{\beta}{L\beta_0};\,\,\tau_2\equiv\frac{\beta}{La_0},
\end{eqnarray}
defining the complex number $\tau$ in Eq.~(\ref{tau_torus}).
Without changing the low-energy physics (e.g., long-distance correlations),
we can take the critical parafermionic chain $\mathcal{H}_f$ in the thermodynamic limit $L\rightarrow\infty$ ($\xi\sim La_0$ and $\xi/a_0\rightarrow\infty$), 
to be at the corresponding infrared renormalization-group (RG) fixed point.
Its bosonic dual $H_b$ is also at  the critical RG fixed point, since $\mathcal{H}_f$ and $H_b$ are locally indistinguishable in the thermodynamic limit, where the effect of different choices of boundary conditions disappears. 
Up to some non-universal factor~\cite{Sachdev:2007aa}, the critical bosonic partition function $Z^\text{latt}_{a_1,a_2}$ in Eq.~(\ref{mod_partition_b}) converges to that of CFTs in the continuum limit as
\begin{eqnarray}
Z_{a_1,a_2}(\tau)&\equiv&
\text{Tr}\!\left[(Q_b)^{ a_2 }q^{L^b_0(a_1)-c/24}\bar{q}^{\bar{L}^b_0(a_1)-c/24}\right],
\end{eqnarray}
where $q\equiv\exp(2\pi i\tau)$ and $\bar{q}\equiv q^*$ with $\tau\equiv\tau_1+i\tau_2$, and $L_0^{b}(a_1)$ and $\bar{L}_0^{b}(a_1)$ are $a_1$-twisted conformal-transformation generators of the bosonic CFT with a central charge $c$~\cite{Francesco:2012aa}.
It follows from Eq.~(\ref{para_00}) that
the parafermionic partition function $\mathcal{Z}^\text{latt}_{s_1,s_2}$ in Eq.~(\ref{mod_partition_f}) converges to
\begin{eqnarray}
\label{parafermionization_1}
\mathcal{Z}_{s_1,s_2}(\tau)&=&\frac{1}{k}\sum_{a_1,a_2}z^{s_1,s_2}_{-a_1,-a_2}\,Z_{a_1,a_2}(\tau)
\end{eqnarray}
in the continuum limit.
Indeed, the underlying lattice system with the partition function $\mathcal{Z}_{s_1,s_2}(\tau)$ is at a critical point since it depends only on the dimensionless ratio $\tau_2=\beta/La_0$ or $\tau_1=(\beta/L\beta_0)$ rather than any length scale. 
In addition, we know that the partition function of a critical bosonic system as a CFT, in the absence of $\mathbb{Z}_k$ or gravitational anomaly, obeys the modular-invariance condition~\footnote{In this work, we use ``modular invariance'' to denote the invariance of physics under the modular transformations of the spacetime torus. When describing the partition functions, it should be implicitly understood as ``modular covariance''.} due to the emergent large-diffeomorphism invariance in conformal invariant field theories~\cite{Polchinski:1998aa}:
\begin{subequations} 
\label{modular invariance boson}
\begin{eqnarray}
Z_{a_1,a_2+a_1}(\tau+1)&=&Z_{a_1,a_2}(\tau);\\
Z_{-a_2,a_1}(-1/\tau)&=&Z_{a_1,a_2}(\tau).
\end{eqnarray}
\end{subequations}
By the help of the invertibility (\ref{invertible_1}), 
we derive the modular transformation of the partition function of the critical parafermionic chain as
\begin{subequations}
\label{modular}
\begin{eqnarray}
\mathcal{Z}_{s_1,s_2}(\tau+1)&=&\sum_{s_1',s_2'}\mathcal{T}_{s_1,s_2}^{s'_1,s'_2}\,\,\mathcal{Z}_{s'_1,s'_2}(\tau);\\
\mathcal{Z}_{s_1,s_2}(-1/\tau)&=&\sum_{s_1',s_2'}\mathcal{S}_{s_1,s_2}^{s'_1,s'_2}\,\,\mathcal{Z}_{s'_1,s'_2}(\tau), 
\end{eqnarray}
\end{subequations}
where
\begin{subequations}
\label{T & S}
\begin{eqnarray}\label{T}
\mathcal{T}_{s_1,s_2}^{s'_1,s'_2}&=&\frac{1}{k^2}\sum_{a_1,a_2}z^{s_1,s_2}_{a_1,a_2+a_1}\left(z^{s_1',s_2'}_{a_1,a_2}\right)^*;\\
\label{S}
\mathcal{S}_{s_1,s_2}^{s'_1,s'_2}&=&\frac{1}{k^2}\sum_{a_1,a_2}z^{s_1,s_2}_{-a_2,a_1}\left(z^{s_1',s_2'}_{a_1,a_2}\right)^*.
\end{eqnarray}
\end{subequations}
These two transformations reduce to 
\begin{subequations}
\label{gauge_2}
\begin{eqnarray}
\mathcal{Z}_{s_1,s_2+s_1}(\tau+1)|_{k=2}&=&\mathcal{Z}_{s_1,s_2}(\tau)|_{k=2};\\
\mathcal{Z}_{-s_2,s_1}(-1/\tau)|_{k=2}&=&\mathcal{Z}_{s_1,s_2}(\tau)|_{k=2}
\end{eqnarray}
\end{subequations}
in the case of fermionization when $k=2$, in agreement with the bosonic case (\ref{modular invariance boson}).
Thus the fermionic spin structure appears as a $\mathbb{Z}_2$-gauge field on the torus. 
However, for general $k>2$, the paraspin structure no longer behaves as Eq.~(\ref{gauge_2}) under the modular transformation although it plays a similar role of $\mathbb{Z}_k$ twisting on a torus. 

\begin{figure}
\begin{center}
\includegraphics[width=12cm,pagebox=cropbox,clip]{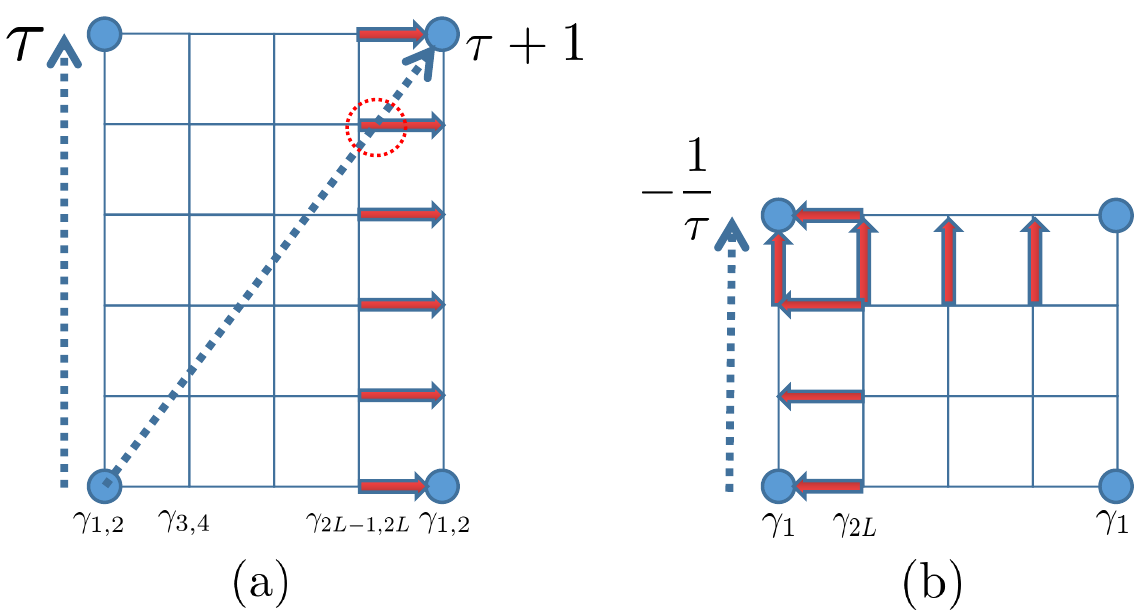}
\caption{(a) $\mathcal{T}$ transformation of spacetime torus: the transformation $\tau\rightarrow\tau+1$ crosses an additional branch cut during the continuum limit $a_0,\beta_0\rightarrow0$. (b) $\mathcal{S}$ transformation: the space-time plane is effectively rotated by 90 degrees, after which the branch-cutting link along the spatial direction is reversed.}
\label{T_S_transformation}
\end{center}
\end{figure}

To understand this unconventional transformation when $k>2$, 
let us consider, for simplicity, $\tau\in i\mathbb{R}$ as in FIG.~\ref{T_S_transformation}.
We start from a $\mathcal{T}$ transformation, $\tau\rightarrow\tau+1$. 
In addition to the $\mathbb{Z}_k$ twisting, 
the solid arrows in FIG.~\ref{T_S_transformation}~(a) also carry the algebraic information in that they connect the last site $2L$ and the ``next'' site $1$, which, according to the algebra in Eq.~(\ref{para_1}), in a distinct way from the other links connecting two neighboring $\gamma_j$ and $\gamma_{j+1}$ when $\omega\neq\omega^*$.
Such a piece of algebraic information is irrelevant or invisible when $k=2$ because $\omega^*|_{k=2}=\omega|_{k=2}$.
When $k>2$, the partition function on the $\mathcal{T}$-transformed space-time torus cannot be identified with $\mathcal{Z}_{s_1,s_2+s_1}(\tau+1)$ due to the additional appearance of the algebraic change on the link circled in FIG.~\ref{T_S_transformation}~(a), crossed by the temporal arrow during the continuum limit $a_0,\beta_0\rightarrow0$, although the $s_1$-twisting information on that link can be fused with $s_2$ to be $(s_2+s_1)$ by a temporal $\mathbb{Z}_k$-gauge transformation. 
For an $\mathcal{S}$ transformation $\tau\rightarrow-1/\tau$,
we assume that there are also solid arrows along the temporal direction to represent similar branch cuts so that the space and time are on the same footing in order to have a $\mathcal{S}$-transformation property
since $\mathcal{S}$ transformations partially have the effect of interchanging space and time.
However, no matter how these temporal arrows are oriented,
either the spatial or temporal orientation will be reversed by the $\mathcal{S}$ transformation as in FIG.~\ref{T_S_transformation}~(b), for example.
That the orientation is relevant when $k>2$
makes the current $\mathcal{S}$-transformation rule (\ref{S}) unconventional as well.
On the other hand, 
the cases of $k=2$ do not have these problems since the orientation of the closing link is irrelevant due to $\omega|_{k=2}=-1=\omega^*|_{k=2}$ there. 
We will also see in the next section that no matter how we adjust the reference ``periodic'' point of $s_{1,2}=0$, the modular transformation~(\ref{modular}) cannot be the same as the conventional form when $k>2$.

Furthermore, since we consider general critical parafermions without reference to Hamiltonian, 
we conclude that the modular transformations~(\ref{modular}) and (\ref{T & S}) as the modular invariance requirement and consistency conditions for parafermionic systems~(\ref{para_1}) at criticality.

\section{Traditional $\mathbb{Z}_k$-paraspin invertible topological phases}\label{invertible}
In this section, we will show that if the partition function $z^{s_1,s_2}$ of an invertible topological field theory with a $\mathbb{Z}_k$-paraspin structure obeyed the traditional modular invariance by $\mathcal{T}$ and $\mathcal{S}$,
\begin{eqnarray}\label{traditional}
z^{s_1,s_2+s_1}=z^{s_1,s_2};\,\,z^{-s_2,s_1}=z^{s_1,s_2},
\end{eqnarray}
then it is always trivial (equals to $1$) when $k\in2\mathbb{Z}+1$, or behaves as an $\mathbb{Z}_2$-Arf invariant,
\begin{eqnarray}
z^{s_1,s_2}=(-1)^{(1+s_1)(1+s_2)},
\end{eqnarray}
in addition to the trivial phase on a torus when $k\in2\mathbb{Z}$, 
contradicting our result of Eq.~(\ref{tft}). 
Furthermore, the partition function $\mathcal{Z}_{s_1,s_2}$ would transform conventionally as Eq.~(\ref{gauge_2}) if the attached topological phase $z^{s_1,s_2}$ on the bosonic theory $Z_{a_1,a_2}$ obeys the modular invariance~(\ref{traditional}).
In this sense, 
the traditional modular invariance is inapplicable to our parafermionic chains.

We start with a definition of modular-transformation orbits: 
\begin{eqnarray}\label{mod}
[(s_1,s_2)]\equiv\left\{\left(\begin{array}{cc}a&b\\c&d\end{array}\right)\!\!\left(\begin{array}{c}s_1\\s_2\end{array}\right):\left(\begin{array}{cc}a&b\\c&d\end{array}\right)\in SL(2,\mathbb{Z})\right\},
\end{eqnarray}
in addition to the $\mathbb{Z}_k$ property of the paraspin structure: 
\begin{eqnarray}\label{z_k}
[(s_1,s_2)]\equiv [(s_1+k,s_2)]\equiv[(s_1,s_2+k)].
\end{eqnarray}
We first prove that
\begin{eqnarray}\label{orbits}
[(s_1,s_2)]=[\text{gcd}(s_1,s_2,k),0]=[0,\text{gcd}(s_1,s_2,k)],
\end{eqnarray}
where ``gcd'' denotes the (non-negative) greatest common divisor. 
It can be shown as follows. 
We note that
\begin{eqnarray}
\text{gcd}\left[\frac{s_1}{\text{gcd}(s_1,s_2,k)},\frac{s_2}{\text{gcd}(s_1,s_2,k)},\frac{k}{\text{gcd}(s_1,s_2,k)}\right]=1. 
\end{eqnarray}
Then, by Dirichlet's theorem on arithmetic progressions, 
there exists an integer $N>k$ such that
\begin{eqnarray}
N\frac{s_1}{\text{gcd}(s_1,s_2,k)}+\frac{s_2}{\text{gcd}(s_1,s_2,k)}
\end{eqnarray}
is a prime number thereby
\begin{eqnarray}
&&\text{gcd}\left[N\frac{s_1}{\text{gcd}(s_1,s_2,k)}+\frac{s_2}{\text{gcd}(s_1,s_2,k)},\frac{k}{\text{gcd}(s_1,s_2,k)}\right]=1\nonumber
\end{eqnarray}
and then
\begin{eqnarray}
&&\text{gcd}\left(N{s_1}{}+{s_2}{},{k}{}\right)=\text{gcd}(s_1,s_2,k). 
\end{eqnarray}
It implies that there exists a unique integer $M$ so that
\begin{eqnarray}\label{master}
M(Ns_1+s_2)=\text{gcd}(s_1,s_2,k)\mod k. 
\end{eqnarray}
By Eqs.~(\ref{z_k}, \ref{mod}, \ref{master}), 
we obtain
\begin{eqnarray}
[(Ns_1+s_2,-s_1)]&=&[(Ns_1+s_2,\text{gcd}(s_1,s_2,k))]\nonumber\\
&=&[(\text{gcd}(s_1,s_2,k),-Ns_1-s_2)]\nonumber\\
&=&[\text{gcd}(s_1,s_2,k),0].
\end{eqnarray}
On the other hand,
we have the following equivalence relation within the orbit
\begin{eqnarray}
[(s_1,s_2)]=[(s_1,Ns_1+s_2)]=[(Ns_1+s_2,-s_1)]
\end{eqnarray}
hence giving Eq.~(\ref{orbits}). 
Thus, any $(s_1,s_2)$ must fall in a class $[(r,0)]=[(0,r)]$ where $r$ divides $k$.
In addition, $[(r,0)]\neq [(r',0)]$ if both $r$ and $r'$ divide $k$ and $r\neq r'$,
because $[(r,0)]$ only contains the elements of the form as $(r\mathbb{Z},r\mathbb{Z})$. 

In a short summary, 
the modular invariance requires the following form on the partition function of the topological field theory $z^{s_1,s_2}$: 
\begin{eqnarray}
z^{s_1,s_2}=f\bigl(\text{gcd}(s_1,s_2,k)\bigr). 
\end{eqnarray}

Since the partition function is defined up to a $\mathbb{Z}_k$ phase,
let us impose a normalization: 
\begin{eqnarray}
f(1)=1, 
\end{eqnarray}
which implies $z^{1,0}=z^{0,-1}=1$. 
Therefore, the value of $z^{0,0}$ exactly gives the $\mathbb{Z}_k$ charge of the untwisted sector. 

For general $k$, we know that the ratio $z(0,1)/z(0,-1)=1$ is the square of $\mathbb{Z}_k$ charge of the untwisted sector by the operator formalism of the partition function.
The normalization above implies that $z^{(0,0)}=\pm1$ and
\begin{eqnarray}
\frac{z^{(0,s_2)}}{z^{(0,-1)}}=\left(z^{(0,0)}\right)^{s_2+1}=(\pm1)^{s_2+1}.
\end{eqnarray}

When $k\in2\mathbb{Z}+1$, then $[(0,1)]=[(0,2)]$ which implies that the untwisted sector has a trivial $\mathbb{Z}_k$ charge, namely $z^{(0,0)}=1$.
Then, $z^{(s_1,s_2)}=1$ for any $(s_1,s_2)$ by modular invariances. 

When $k\in2\mathbb{Z}$, we can have two phases: $z^{(s_1,s_2)}=1$ and a nontrivial one satisfying
\begin{eqnarray}
z^{(0,s_2)}=(-1)^{s_2+1}, 
\end{eqnarray}
which gives the result as
\begin{eqnarray}
z^{(s_1,s_2)}=(-1)^{(1+s_1)(1+s_2)}, 
\end{eqnarray}
reduced to the Arf invariant when treating $s_{1,2}$ as modulo $2$.
In conclusion, we could have one single $\mathbb{Z}_2$ nontrivial phase only if $k\in2\mathbb{Z}$. 

Therefore, when $k>2$ especially $k\in2\mathbb{Z}+1$, traditionally modular invariant $\mathbb{Z}_{k}$-paraspin topological field theories~(\ref{traditional}) cannot correctly describe the low-energy property of gapped $\mathbb{Z}_{k}$-parafermionic chains
and modular invariance for parafermionic systems should be modified to be Eq.~(\ref{modular}). 

\section{Minimal parafermionic models and fractional statistics}\label{fractional_stat}
Minimal models,
or more precisely, Virasoro minimal models are an essential concept in two-dimensional CFTs, whose energy spectra can be arranged into finitely many irreducible representations of the Virasoro algebra.
The bosonic minimal models have been exactly solved and classified into an ADE classification~\cite{Cappelli:1987aa,Kato:1987aa}.
The fermionic (i.e., $\mathbb{Z}_{k=2}$-parafermionic) minimal models are also exhausted by fermionizations of the bosonic minimal models with a global $\mathbb{Z}_2$ symmetry~\cite{Hsieh:2020aa,Kulp:2020aa,Gaiotto:2020aa}.

In the first part of this section,
we will obtain the remaining minimal $\mathbb{Z}_k-$parafermionic models with $k>2$ by our parafermionization/bosonization method. 

In the second part,
we will derive the fixed-point partition functions of a large class of critical parafermionic chains that are not necessarily Virasoro minimal.

In addition,
it is well known that the fermionization ($k=2$) of the critical quantum transverse Ising model yields the massless Majorana fermions obeying fermion statistics rather than bosonic statistics, 
although the local operators of the Ising model are all bosonic. 
As we will see, 
when $k>2$, the conformal spins of the fundamental field operators in the critical parafermionic theory can be neither integral (bosonic) nor half-integral (fermionic).

\subsection{Minimal $\mathbb{Z}_{k>2}-$parafermionic models}
\label{minimal}
Let us take a $\mathbb{Z}_{k>2}$ parafermionic model, which is Virasoro minimal, 
i.e., its spectrum is built from finitely many irreducible Virasoro representations.
Then, we bosonize it by {our bosonization formula~(\ref{bosonization}) to obtain} a bosonic theory.
Such a bosonic theory must also be Virasoro minimal since the summation in Eq.~(\ref{bosonization}) is finite thereby unchanging the {Virasoro-minimal nature} of the spectrum. 
Moreover, this bosonic theory has a global $\mathbb{Z}_{k>2}$ symmetry ``inherited'' from the parafermionic model. 
However,
it has been shown that the only bosonic minimal models with a global $\mathbb{Z}_{k>2}$ symmetry are the critical and the tricritical three-state Potts models~\cite{Ruelle:1998aa,Kulp:2020aa,Gaiotto:2020aa}, where $k=3$.
The central charges are $c=4/5$ for the critical Potts model and $c=6/7$ for the tricritical Potts model.
Therefore, 
the bosonic theory above can only be one of these two models.
Finally,
we can parafermionize this bosonic theory back by Eq.~(\ref{parafermionization}) to the original $\mathbb{Z}_{k=3}$-parafermionic minimal theory.
We consider these two possibilities below.

\subsubsection{Parafermionic minimal model dual to the critical three-state Potts model $(c=4/5)$}
The first $\mathbb{Z}_{k>2}-$parafermionic minimal model has its bosonization as the critical theory of the three-state Potts model at its ferromagnetic self-dual point. 
Under a twisted boundary condition, the lattice bosonic Hamiltonian takes the form as:
\begin{eqnarray}
{H}^{a_1}_{\text{Potts}}=-\sum_{p=1}^{2}\left\{\sum_{j=1}^{L-1}
\left[(\sigma_j)^p(\sigma_{j+1})^{3-p}+(\tau_j)^p\right]
+(\sigma_L)^p\left(\sigma_{1}\omega^{a_1}\right)^{3-p}+(\tau_L)^p\right\}.
\end{eqnarray} 
Its (fixed-point) critical partition function under the periodic boundary condition is
\begin{eqnarray}\label{Potts_univ}
Z^{0,0}_\text{Potts}=|\chi_0+\chi_3|^2+|\chi_{2/5}+\chi_{7/5}|^2+2|\chi_{2/3}|^2+2|\chi_{1/15}|^2, 
\end{eqnarray}
where $\chi_{h}$ is the character of an irreducible representation of Virasoro algebra with a conformal dimension $h$ and its form under general twisting can be seen later in Eq.~(\ref{self}) (where $c^0_0=\chi_0+\chi_3$, $c_2^0=\chi_{2/3}$, $c^1_1=\chi_{1/15}$ and $c^2_0=\chi_{2/5}+\chi_{7/5}$). 
Although the complete (chiral) primary operator content of the Potts model contains parafermionic operators, they do not enter into the local-operator content determining the partition function $Z^{0,0}_\text{Potts}$, namely any character product $\chi_{h_i}{\chi}^*_{h_j}$ in the expansion of $Z^{0,0}_\text{Potts}$ above satisfies the bosonic statistics: 
\begin{eqnarray}
h_i-h_j\in\mathbb{Z}. 
\end{eqnarray}
It implies the Potts model is bosonic in nature, as it should be.

Then we can evaluate the partition function $\mathcal{Z}_{k=3}^{s_1,s_2}$ of the parafermionic minimal model, 
e.g., with the lattice realization 
\begin{eqnarray}
\mathcal{H}_{k=3}^{s_1}=\sum_{j=1}^{2L-1}\left(-\omega\gamma_{j}^\dagger\gamma_{j+1}\right)-\omega^{1+s_1}\gamma_{2L}^\dagger\gamma_1+\text{h.c.},
\end{eqnarray}
whose bosonic correspondence by Eq.~(\ref{bosonization}) is in the same universality class as the Potts {model~(\ref{Potts_univ})}.
We select the paraspin structure as $(1+s_1,1+s_2)=(0,0)$ so that the ground state of so-twisted Hamiltonians can be generated by a partition function on the complex plane without any operator insertion other than the identity operator,
i.e., the partition function on torus having $\chi_0\chi^*_0$.
The reason for such a choice is discussed {in detail} in Appendix~\ref{locality}, which turns out to be related to the conformal spins of parafermionic fields.
For the paraspin structure $(1+s_1, 1+s_2)=(0,0)$,
by Eq.~(\ref{parafermionization_1}),
we have
\begin{eqnarray}
\mathcal{Z}^{-1,-1}_{k=3}=|\chi_0+\chi_3|^2+|\chi_{2/5}+\chi_{7/5}|^2+2(\chi_0+\chi_3)\chi_{2/3}^*+2(\chi_{2/5}+\chi_{7/5})\chi_{1/15}^*, 
\end{eqnarray}
of which several operators have fractional statistics, i.e., other than bosonic or fermionic statistics. 
They can be seen more clearly when represented by the scaling dimension $\Delta$ and conformal spin $\mathcal{S}$: 
\begin{eqnarray}
\Phi_{\Delta,\mathcal{S}}\equiv\chi_{(\Delta+\mathcal{S})/2}\chi^*_{(\Delta-\mathcal{S})/2}, 
\end{eqnarray}
which gives
\begin{eqnarray}\label{state-operator}
\mathcal{Z}^{-1,-1}_{k=3}&=&\Phi_{0,0}+\Phi_{3,3}+\Phi_{0,3}+\Phi_{3,0}+\Phi_{4/5,0}+\Phi_{14/5,0}+\Phi_{9/5,-1}+\Phi_{9/5,1}\nonumber\\
&&+2\Phi_{2/3,-2/3}+2\Phi_{11/3,7/3}+2\Phi_{7/15,1/3}+2\Phi_{22/15,4/3}.
\end{eqnarray}
Interestingly, all the operators above have spins that are multiples of $1/3$, i.e., $\mathcal{S}\in\mathbb{Z}/3$, 
while the characters like $\chi_{2/3}\chi^*_{7/5}$ with anomalous spins are not allowed to occur in the partition function. 
It is consistent with the fact that the fundamental degrees of freedom are $\mathbb{Z}_3$ parafermions with the spins $\pm1/3$ as the elementary units.
The partition functions under other paraspin structures are given in Appendix~\ref{calculation}, which complete the first one of the two $\mathbb{Z}_{k>2}$-parafermionic minimal models.

Let us discuss more on the spectra~(\ref{state-operator}).
Besides the bosonic operators like the identity operator $\Phi_{0,0}$ and the energy operator $\Phi_{3,3}$, the operator content also contains the fundamental parafermionic field $\Phi_{2/3,-2/3}$, which is genuinely parafermionic with fractional statistics.
Furthermore, this parafermion operator $\Phi_{2/3,-2/3}$ is even not mutually local with itself due to {its} multivalued two-point correlator calculated in traditional CFTs~\cite{Zamolodchikov:1985aa,Gepner:1987aa}.
In addition, 
the field of $\Phi_{2/3,2/3}$ which is mutually local (by the calculations in traditional CFTs) with the existing $\Phi_{2/3,-2/3}$ does not show up above.
However, the state-operator correspondence (see Appendix~\ref{locality}) of traditional bosonic/fermionic CFTs implies that all the operators in Eq.~(\ref{state-operator}) are local/mutually local~\cite{Polchinski:1998aa}.
Therefore, 
the concept of (mutual) locality of the critical parafermionic field theories is sharply distinct from the bosonic/fermionic CFTs.

\subsubsection{Parafermionic minimal model dual to the tricritical three-state Potts model $(c=6/7)$}
The {other} parafermionic minimal model has its bosonization as the tricritical three-state Potts model.
The bosonic theory has the partition functions under boundary twistings as~\cite{Zuber:1986aa,Ruelle:1998aa}:
\begin{eqnarray}
\label{tripotts_0}
Z_\text{tr-Potts}^{0,0}&=&|\chi_0+\chi_5|^2+|\chi_{1/7}+\chi_{22/7}|^2+|\chi_{5/7}+\chi_{12/7}|^2\nonumber\\
&&+2|\chi_{4/3}|^2+2|\chi_{10/21}|^2+2|\chi_{1/21}|^2,\\
\label{tripotts_1}
Z_\text{tr-Potts}^{0,1}=Z_\text{tr-Potts}^{0,2}&=&|\chi_0+\chi_5|^2+|\chi_{1/7}+\chi_{22/7}|^2+|\chi_{5/7}+\chi_{12/7}|^2\nonumber\\
&&+(\omega+\omega^2)\left(|\chi_{4/3}|^2+|\chi_{10/21}|^2+|\chi_{1/21}|^2\right),
\end{eqnarray}
and
\begin{eqnarray}
\label{tripotts_2}
&&Z_\text{tr-Potts}^{1,a_2}=Z_\text{tr-Potts}^{2,-a_2}\nonumber\\
&=&\omega^{2a_2}\left[\chi_{4/3}(\chi_0+\chi_5)^*+\chi_{10/21}(\chi_{1/7}+\chi_{22/7})^*+\chi_{1/21}(\chi_{5/7}+\chi_{12/7})^*\right]+\text{c.c.}\nonumber\\
&&+|\chi_{4/3}|^2+|\chi_{10/21}|^2+|\chi_{1/21}|^2.
\end{eqnarray}
Then,
we parafermionize it by Eq.~(\ref{parafermionization_1}) back to the parafermionic minimal model with the paraspin $(s_1,s_2)$ satisfying $(1+s_1,1+s_2)=(0,0)$ to see the fundamental operator content as explained in Appendix~\ref{locality}:
\begin{eqnarray}
{\mathcal{Z}'_3}^{-1,-1}&=&|\chi_0+\chi_5|^2+|\chi_{1/7}+\chi_{22/7}|^2+|\chi_{5/7}+\chi_{12/7}|^2\nonumber\\
&+&2\left[\chi_{4/3}(\chi_0+\chi_5)^*+\chi_{10/21}(\chi_{1/7}+\chi_{22/7})^*+\chi_{1/21}(\chi_{5/7}+\chi_{12/7})^*\right],
\end{eqnarray}
or expressed by scaling dimensions and conformal spins:
\begin{eqnarray}
{\mathcal{Z}_3'}^{-1,-1}&=&\Phi_{0,0}+\Phi_{5,-5}+\Phi_{5,5}+\Phi_{10,0}+\Phi_{2/7,0}+\Phi_{23/7,-3}+\Phi_{23/7,3}+\Phi_{44/7,0}+\Phi_{10/7,0}\nonumber\\
&&+\Phi_{17/7,-1}+\Phi_{17/7,1}+\Phi_{24/7,0}+2\left(\Phi_{4/3,4/3}+\Phi_{19/3,-11/3}+\Phi_{13/21,1/3}\right.\nonumber\\
&&\left.+\Phi_{76/21,-8/3}+\Phi_{16/21,-2/3}+\Phi_{37/21,-5/3}\right),
\end{eqnarray}
from which the conformal spins of the operator content can be also shown to be consistently the multiples of $1/3$, i.e., $\mathcal{S}\in\mathbb{Z}/3$.
The partition functions under other paraspin structures $(s_1,s_2)$ can be found in Appendix~\ref{calculation}, which complete the derivation of the last $\mathbb{Z}_{k}$-parafermionic minimal model. 

So far, we have exhausted all the $\mathbb{Z}_{k}$-parafermionic minimal models in addition to the completed fermionic minimal models~\cite{Hsieh:2020aa,Kulp:2020aa}.
In the following part,
we will consider a large class of parafermionic models which are not necessarily Virasoro minimal.

\subsection{Critical parafermionic theories dual to $\mathbb{Z}_k$-clock models}
Critical $\mathbb{Z}_k$-clock models with general interactions are expected to realize the following partition function with the central charge $c=2(k-1)/(k+2)$~\cite{Zamolodchikov:1985aa,Gepner:1987aa}:
\begin{eqnarray}\label{self}
Z_{k}^{a_1,a_2}=\sum_{m=-k+1}^k\sum_{l,\bar{l}}\frac{1}{2}|\eta(\tau)|^2\omega^{-(m+a_1)a_2}L_{l,\bar{l}}c^l_m(\tau)c^{\bar{l}*}_{m+2a_1}(\tau),
\end{eqnarray}
where $\eta(\tau)$ is the Dedekind's $\eta$ function and the integral multiplicity $L_{l,\bar{l}}$ $(0\leq l,\bar{l}\leq k)$ is given by the modular invariant solutions $L_{l,\bar{l}}\chi_{h_{l}}\chi^*_{h_{\bar{l}}}$ of the SU$(2)$-current algebra system~\cite{Francesco:2012aa}.
Here, $c^l_m(\tau)$'s are string functions derivable from the coset $\hat{su}(2)_k/\hat{u}(1)$ construction satisfying $c^l_m=c^l_{m+2k}=c^l_{-m}=c^{k-l}_{k+m}$ with $c^l_{m\neq l\text{ mod }2}\equiv0$~\cite{Francesco:2012aa}, and $|\eta(\tau)|^2c_{m}^l(\tau)c_{{m}-2a_1}^{\bar{l}*}(\tau)$ can be seen as the partition function of the primary field with the $\mathbb{Z}_k\times\hat{\mathbb{Z}}_k$ charge of $(m-a_1,0)$.
The self-duality can be proven by the invariance under $\mathbb{Z}_k$ gauging, which generalizes the Kramers-Wannier duality~\cite{Kramers:1941aa,Kramers:1941ab,Kogut:1979aa} in that gauging $\mathbb{Z}_k$ effectively neutralizes the $\mathbb{Z}_k$ charge or eliminates the nontrivially $\mathbb{Z}_k$-charged primary fields, e.g., spin order parameters, from the local-operator spectrum which thus only contains purely $\hat{\mathbb{Z}}_k$-charged fields, i.e., various disorder parameters.  
The parafermionization of Eq.~(\ref{self}) by Eq.~(\ref{parafermionization_1}) yields the partition function in the universality class of a large class of critical $\mathbb{Z}_k$-parafermionic chains,
\begin{eqnarray}\label{parafermionization_2}
\mathcal{Z}_{k}^{s_1,s_2}&=&\sum_{l,\bar{l}}\sum_{a_1=0}^{k-1}\frac{\omega^{(1+s_1-a_1)(1+s_2)}}{2/|\eta(\tau)|^2}L_{l,\bar{l}}\left[c^{l}_{1+s_1}(\tau)c_{1+s_1-2a_1}^{\bar{l}*}(\tau)+c^{l}_{1+s_1+k}(\tau)c_{1+s_1+k-2a_1}^{\bar{l}*}(\tau)\right], \nonumber\\
\end{eqnarray}
which explicitly satisfies the consistency condition of the unconventional modular invariance~(\ref{T}, \ref{S}). 
The conformal spins of the local physical operator content of the $s_1$-twisted CFT with $1+s_1=0$ are generically fractional and can be read from Eq.~(\ref{parafermionization_2}) with $s_1=s_2=-1$ (see Appendix~\ref{locality}) as $\{\mathcal{S}\}_{s_1=-1}$ of
\begin{eqnarray}
\{\mathcal{S}\}_{s_1}=\left\{h^l_{1+s_1+\lambda k}-h^{\bar{l}}_{1+s_1-2a+\lambda k}\left|L^{l,\bar{l}}\neq0,\,\lambda=0,1\text{ and }a=1,\cdots,k\right.\right\},
\end{eqnarray}
where $h^l_m$ is the conformal dimension related to $c^l_m(\tau)$~~\cite{Zamolodchikov:1985aa,Gepner:1987aa}: 
\begin{eqnarray}
h^l_m=\left\{\begin{array}{ll}\frac{l(k-l)}{2k(k+2)}+\frac{(l-m)(l+m)}{4k},&-l\leq m\leq l;\\
\frac{l(k-l)}{2k(k+2)}+\frac{(l-m)(l+m-2k)}{4k},&l< m<2k-l,
\end{array}\right.
\end{eqnarray}
in which $m$ is set to stay in the interval $[-l,2k-l)$ by the cyclicity $c^l_m=c^l_{m+2k}$ given before.

\section{Conclusions}\label{conclusion}
In this work,
we propose a one-dimensional $\mathbb{Z}_k$-parafermionization/bosonization scheme on critical parafermionic chains starting from a generalized Jordan-Wigner transformation.
It is shown to be equivalent to an attachment construction of attaching a nontrivial topological phase of a gapped parafermionic chain, generalizing the conventional fermionization/bosonization. 
Such a parafermionization enables us to study the critical parafermionic system whose fundamental degrees of freedom are fractionally statistical fields beyond bosons and fermions. 
We find that the critical theories of the parafermions generally obey unconventional modular-transformation rules and potentially have distinct concepts of mutual locality, which are not in the framework of existing bosonic/fermionic CFTs.
Such a modular invariance requirement can be taken as general consistency conditions for parafermionic critical field theories realizable on lattices.
Its implication on the anomaly of parafermionic field theories and the concept of mutual locality in parafermions when $k>2$ can be of future interest.
We also apply our result to exhaust all the $\mathbb{Z}_k$-parafermionic minimal models together with earlier works on $k=2$. {The applications to} rational parafermionic models with more complicated symmetries are expected in future works.

\acknowledgments
The authors are grateful to Hosho Katsura and Yuji Tachikawa for helpful advice on the manuscript.
Y.~Y. was supported by JSPS fellowship.
This work was supported in part by JSPS KAKENHI [Grants No.~JP19J13783 (Y.Y.) and No.~19K03680 (A.F.)] and by JST CREST [Grant No.~JPMJCR19T2 (A.F.)].

\appendix
\section{Space-time torus on lattices and modular invariance}
\label{lattice_torus}
In this part, 
we review how to formulate general $(1+1)$-dimensional bosonic CFTs twisted by an onsite abelian symmetry $G$, taken as $G=\mathbb{Z}_k$ for simplicity, on a general space-time torus parametrized by a complex number $\tau\equiv\tau_1+i\tau_2$, starting from the underlying lattice model. 

Let us consider a bosonic chain with the bosonic degrees of freedom $\sigma_i$ at the site $i$ in short, 
e.g., $\sigma_i$ being $(\sigma^x_i,\sigma^z_i)$ for the Ising model.
The system is of length $L$ and we identify
\begin{eqnarray}
\sigma_{L+1}\equiv\omega^r\sigma_1.
\end{eqnarray}
The Hamiltonian is twisted at the link connecting $\sigma_{L}$ and $\sigma_1$ by a symmetry operation $r$~mod~$k\in\mathbb{Z}_k$ as in Eq.~(\ref{twisted bc for boson}), and {the twisted Hamiltonian} is denoted as $H_b(r)$. 
Additionally,
we also denote the $\mathbb{Z}_k$ generator as $Q(\sigma_1,\sigma_2,\cdots,\sigma_L)$ which is $r$-independent. 

\begin{figure}
\begin{center}
\includegraphics[width=9cm,pagebox=cropbox,clip]{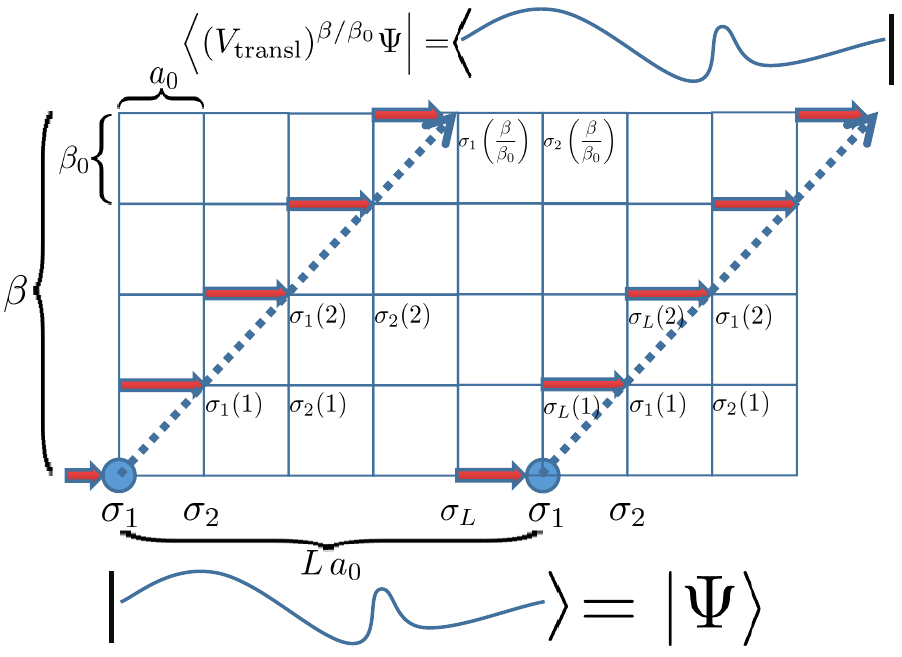}
\caption{The visualization of the partition function $Z^\text{latt}_{r,s}$: the dashed arrows point along the vector $(\tau_1,\tau_2)$ and the state $|\Psi\rangle$ is summed over.
The solid arrows denote the $r$-twisted link between the sites $L$ and $1$ and the correspondng links moved by $V_\text{transl}$ during the time evolution.
The temporal $s$-twisting is not shown explicitly in this figure.}
\label{visual}
\end{center}
\end{figure}

For later convenience, 
we define the common translation transformation:
\begin{eqnarray}
V_\text{transl}\sigma_iV_\text{transl}^\dagger=\sigma_{i+1}.
\end{eqnarray}
We consider the following imaginary-time evolution by the time-dependent Hamiltonian during the time $[0,(\beta/\beta_0)\beta_0]$: 
\begin{eqnarray}
H_b(t;r)=\left(V_\text{transl}\right)^nH_b(r)\left(V_\text{transl}^\dagger\right)^n,\text{ when }t\in[n\beta_0,(n+1)\beta_0).
\end{eqnarray}
Here $\beta_0$ can be seen as a ``lattice spacing" along the imaginary time and $a_0$ is the (spatial) lattice spacing.
Similarly,
we can also have a time-dependent $\mathbb{Z}_k$ generator:
\begin{eqnarray}
Q_b(t)=\left(V_\text{transl}\right)^nQ_b\left(V_\text{transl}^\dagger\right)^n,\text{ when }t\in[n\beta_0,(n+1)\beta_0), 
\end{eqnarray}
which is actually time-independent for generic bosonic systems since $\sigma$'s commute between distinct sites.
Then we define the following partition function:
\begin{eqnarray}\label{mod_partition}
Z^\text{latt}_{r,s}\equiv\text{Tr}\left\{(V^\dagger_\text{transl})^{\beta/\beta_0}[Q_b(\beta)]^s\,\,\mathcal{T}\exp\left[-\int_0^\beta dtH_b(t;r)\right]\right\},
\end{eqnarray}
where $\mathcal{T}$ is the time-ordering operator.
Such an evolution and the corresponding partition function $Z^\text{latt}_{r,s}$ can be visualized on the discrete space-time lattice as in FIG.~\ref{visual}.
If the lattice model $H_b(\sigma_1,\sigma_2,\cdots)$ on an infinite chain is at criticality, i.e., if the correlation length diverges $\xi/a_0~\rightarrow~\infty$ in the unit of the lattice constant $a_0$ as the system length $L\rightarrow\infty$,
then its universal properties are described by its lattice RG fixed point.
Then, without changing the low-energy physics, e.g., long-distance correlations, 
we take our lattice Hamiltonian $H_b(\sigma_1,\sigma_2,\cdots)$ to be at the corresponding infrared RG fixed point,
which is described by a CFT after the following continuum limit is taken:
\begin{eqnarray}
a_0\rightarrow0,\text{ while }\beta_0/a_0,\,\beta\text{ and }La_0\text{ fixed},
\end{eqnarray}
and various coupling constants in the lattice Hamiltonian are scaled to keep the correlation length $\xi\sim La_0$ fixed~\footnote{Here, we have already taken the lattice Hamiltonian $H_b$ at its RG fixed point, so the coupling terms do not need to be scaled.}.
We define the following ratios: 
\begin{eqnarray}
\frac{\beta}{La_0}=\tau_2;\,\frac{\beta_0}{a_0}=\frac{\tau_2}{\tau_1},
\end{eqnarray}
which are invariant during the continuum limit.
Then, up to some non-universal factor, the partition function ${Z}^\text{latt}_{r,s}$  converges to the partition function of the corresponding CFT
\begin{eqnarray}\label{mod_cft}
{Z}_{r,s}(\tau)=\text{Tr}\!\left[(Q_b)^sq^{L_0^b(r)-c/24}\bar{q}^{\bar{L}_0^b(r)-c/24}\right],
\end{eqnarray}
where $q\equiv\exp(2\pi i\tau)$ and $\bar{q}\equiv q^*$ with $L_0^{b}(r)$ and $\bar{L}_0^{b}(r)$ the $r$-twisted conformal-transformation generators of the corresponding bosonic CFT with a central charge $c$~\cite{Francesco:2012aa}.
The partition function $Z_{r,s}(\tau)$ has no relevant length scale and it only depends on the dimensionless number $\tau$, which reflects the criticality of the lattice model.

Additionally, in the path-integral formalism of $Z^\text{latt}_{r,s}$ at the fixed point,
the local degrees of freedom $\{\sigma_j\}$ are coarse-grained to be $\phi(t,x)$ with $x=ja_0$, e.g., $\phi(t,x)$ being the configuration of local {spin} density in the Ising model.
The symmetry operation $r\mod k$ acting on $\phi(t,x)$ is denoted as ${}^r\phi(t,x)\equiv\omega^r\phi(t,x)$, and the translation transformation $V_\text{transl}$ {acts on $\phi$} as $|\phi(t,x)\rangle\rightarrow|\phi(t,x-a_0)\rangle$ on the wave functional, where the minus sign should be noted.
The $r$-twisted Hamiltonian and the operators $(V^\dagger_\text{transl})^{\beta/\beta_0}[Q_b(\beta)]^s$ at $t=\beta$ in Eq.~(\ref{mod_partition}) correspond to the following boundary condition in the path integral:
\begin{eqnarray}
{}^r\phi(t,x+La_0)=\phi(t,x);\,\,{}^s\phi\left(t+\beta,x+\frac{\beta}{\beta_0}a_0\right)=\phi(t,x),
\end{eqnarray}
which is translated to, if we define $z=(x+it)/(La_0)$, 
\begin{eqnarray}\label{TBC}
{}^r\phi(z+1)=\phi(z);\,\,{}^s\phi(z+\tau)=\phi(z), 
\end{eqnarray}
as the boundary condition in the path-integral functional integration:
\begin{eqnarray}
Z_{r,s}=\int\mathcal{D}\phi(z)\exp(-S[\phi(z)]),
\end{eqnarray}
reproducing Eq.~(\ref{mod_cft}) for CFTs. 

\section{Relation between local operators and partition functions on a torus}\label{locality}
In this part,
we will explain how to obtain the conformal dimensions of local operators in a CFT from its partition function on a torus by a state-operator correspondence as follows.

Let us consider a CFT with a fundamental field operator $\varphi(z,\bar{z})$ and start from the complex plane $\{z\in\mathbb{C}\}$.
We insert a local operator $\Phi(z,\bar{z})$ made of $\varphi(z,\bar{z})$ at the origin of {the complx plane} and denote the conformal spin of $\varphi(z,\bar{z})$ as $\mathcal{S}_\varphi=h_\varphi-\bar{h}_\varphi$, 
which is integral or half-integral if the CFT is bosonic or fermionic, respectively.
By a path integral,
this insertion defines a quantum state $|\Psi\rangle$ on a unit circle: $S^1\equiv\{z=\exp(i\theta):\theta\in[0,2\pi)\}$ with the wave functional as:
\begin{eqnarray}
\Psi\!\left[\varphi_0\left(z\in S^1\right)\right]=\int_{D^2}\mathcal{D}\varphi\exp\{-S[\varphi]\}\Phi(0,0)|_{\varphi=\varphi_0\text{ on }S^1},
\end{eqnarray}
where $S[\varphi]$ is the action and the path integral is performed on the disk $D^2\equiv\{|z|\leq 1\}$ with the boundary condition $\varphi=\varphi_0$ on $S^1$.
Since $\Phi(0,0)$ is a local operator,
it does not introduce any branch cut for $\varphi$ fields, i.e., PBC still held: 
\begin{eqnarray}\label{pbc_plane}
\varphi\bigl(\exp(2\pi i)z,\exp(-2\pi i)\bar{z}\bigr)=\varphi(z,\bar{z}).
\end{eqnarray}

To evaluate the conformal dimensions $(h_\Phi,\bar{h}_{\Phi})$,
we simply act $L_0$ or $\bar{L}_0$ on $|\Psi\rangle$, for example:
\begin{eqnarray}
\label{dimension}
L_0\Psi\!\!\left[\varphi_0\left(z\in S^1\right)\right]=h_\Phi\Psi\!\!\left[\varphi_0\left(z\in S^1\right)\right].
\end{eqnarray}
Then we apply the following conformal coordinate transformation:
\begin{eqnarray}\label{exp}
z=\exp(w),\,\,\bar{z}=\exp(\bar{w}),
\end{eqnarray}
to transform the theory from $D^2$ to a half-infinitely long cylinder parametrized by $\{w=t+ix|x\sim x+2\pi,t\in(-\infty,0]\}$.
Then, both sides of Eq.~(\ref{dimension}) become:
\begin{eqnarray}
\left(L_0^\text{cyl}+\frac{c}{24}\right)\Psi'[\varphi_0'(ix:x\sim x+2\pi)]=h_\Phi\Psi'[\varphi_0'(ix:x\sim x+2\pi)],
\end{eqnarray}
where $L^\text{cyl}_0=L_0-c/24$ is the generator of Virasoro algebra on the cylinder.
Here $\Psi'$ is the wave functional under the new coordinate on the cylinder and the fundamental field operator is transformed to $\varphi'_0$ by
\begin{eqnarray}
\varphi_0'(w,\bar{w})=\left(\frac{\partial z}{\partial w}\right)^h\left(\frac{\partial \bar{z}}{\partial \bar{w}}\right)^{\bar{h}}\varphi_0(z,\bar{z}),
\end{eqnarray}
which, by Eq.~(\ref{pbc_plane}),
obeys the boundary condition:
\begin{eqnarray}
\exp\left(i2\pi\mathcal{S}_\varphi\right)\varphi'(w+2\pi i,\bar{w}-2\pi i)=\varphi'(w,\bar{w}).
\end{eqnarray}
It implies that the state $|\Psi'\rangle$ is obtained by a path integral on a half-infinitely long cylinder, but with its spatial boundary condition twisted by $\exp(i2\pi\mathcal{S}_\varphi)$.

On the other hand,
the partition function on a torus twisted by $a_1$ in the spatial direction, but without twistings along the temporal direction is 
\begin{eqnarray}
Z_{a_1}\equiv\text{Tr}\exp\left[2\pi\tau \left(L_0^\text{cyl}(a_1)+{\bar{L}}_0^\text{cyl}(a_1)\right)\right],
\end{eqnarray}
for a purely imaginary $\tau\in i\mathbb{R}$.
Then, 
we take $a_1=k\mathcal{S}_\varphi$ when $\exp(i2\pi\mathcal{S}_\varphi)$ twisting is realizable by a $\mathbb{Z}_k$ symmetry, as {is the case with} our paper.
By the definition of the characters $\chi_h$'s,
we obtain that $(h_\Phi,\bar{h}_\Phi)$ must be the conformal dimensions of one of the operators in the conformal family associated with the highest weight $(h,\bar{h})$ appearing in $(A_{h,\bar{h}}\neq0)$
\begin{eqnarray}
Z_{a_1=k\mathcal{S}_\varphi}=\sum_{h,\bar{h}}A_{h,\bar{h}}\chi_h\chi^*_{\bar{h}}.
\end{eqnarray}
Since $\Phi$ is an arbitrary local operator,
the local-operator content of the theory is composed by the operators in the conformal families with the heighest weight $(h,\bar{h})$ for nonzero $A_{h,\bar{h}}$.

\subsection{Examples: Ising model and Majorana fermion ($k=2$)}
Let us consider the Ising model with the central charge $c=1/2$, which is bosonic and the fundamental degree of freedom is the spin operator $\sigma$ whose conformal spin is trivial: $\mathcal{S}_\sigma=0$.
Therefore,
the local-operator content can be directly read from the partition function with an untwisted Hamiltonian: 
\begin{eqnarray}
Z_\text{Ising}^{0,0}=\chi_0\chi_0^*+\chi_{1/2}\chi_{1/2}^*+\chi_{1/16}\chi_{1/16}^*,
\end{eqnarray}
from which we see that the local operators belong to the conformal families of a $(0,0)$ operator (the identity operator), a $(1/2,1/2)$ operator (the energy operator $\epsilon$), and a $(1/16,1/16)$ operator (the spin operator $\sigma$).

For the massless Majorana fermion,
the fundamental degree of freedom is the real fermion $\psi(z)$ with $\mathcal{S}_\psi=-1/2$.
Therefore,
we need to choose
\begin{eqnarray}
s_1=2\mathcal{S}_\psi=-1 \text{ and } 1+s_2=0,
\end{eqnarray}
in the fermionic partition function, 
where the additional ``$1$'' in the ``$1+s_2$'' above is due to our convention $(Q_f)^{1+s_2}$ of the operator insertion along the time direction for the parafermionic partition function~(\ref{para}).
\begin{eqnarray}
\mathcal{Z}_{k=2}^{-1,-1}=\chi_0\chi_0^*+\chi_0\chi_{1/2}^*+\chi_{1/2}\chi_0^*+\chi_{1/2}\chi_{1/2}^*,
\end{eqnarray}
from which we can conclude that the local operators are made from the conformal families of a $(0,0)$ operator (the identity operator), a $(0,1/2)$ operator (the chiral Majorana operator $\psi$), a $(1/2,0)$ operator (the chiral Majorana operator $\bar{\psi}$), and a $(1/2,1/2)$ operator (the mass operator $\bar{\psi}\psi$).

\subsection{Examples: Parafermions with $k>2$}
If we assume that the conditions of the framework developed above, e.g., the conformal transformation~(\ref{exp}), are still applicable for parafermions when $k>2$,
then the fact that the fundamental parafermion field $\psi_{k}(z)$ has a conformal spin $\mathcal{S}_{\psi_k}=-1/k$~\cite{Zamolodchikov:1985aa,Gepner:1987aa} implies that the partition function $\mathcal{Z}_k^{-1,-1}$ can also tell us the local operator content.
However,
as we see in Sec.~\ref{fractional_stat},
the operators in this content can even mutually non-local with themselves.
It means that the notion of locality in parafermionic critical theories is different from traditional bosonic/fermionic CFTs when $k>2$.

\section{Partition functions of the $\mathbb{Z}_{k>2}$ parafermionic minimal models}
\label{calculation}
As argued in Sec.~\ref{minimal},
$\mathbb{Z}_{k>2}$-parafermionic minimal models only exist when $k=3$, and they are dual to the critical and the tricritical three-state Potts models. 
We calculate their partition functions under general paraspin structures below.
 
\subsection{Parafermionic minimal model dual to a critical three-state Potts model}
By the general formalism~(\ref{parafermionization_2}),
 we can calculate the partition function of the parafermionization of ferromagnetic three-state Potts model at its self-dual point, where $L_{l,\bar{l}}=\delta_{l,\bar{l}}$~\cite{Francesco:2012aa}: 
\begin{eqnarray}
\mathcal{Z}^{0,0}_{k=3}=\mathcal{Z}^{1,1}_{k=3}&=&(1+\omega)(|B|^2+|C|^2)+\omega^*(BA^*+CD^*);\\
\mathcal{Z}^{0,1}_{k=3}=\mathcal{Z}^{1,0}_{k=3}&=&(1+\omega^*)(|B|^2+|C|^2)+\omega(BA^*+CD^*);\\
\mathcal{Z}^{0,2}_{k=3}=\mathcal{Z}^{1,2}_{k=3}&=&2(|B|^2+|C|^2)+(CD^*+BA^*); \\
\mathcal{Z}^{2,0}_{k=3}=\mathcal{Z}^{2,1}_{k=3}&=&(|A|^2+|D|^2)+(\omega+\omega^*)(AB^*+DC^*);\\
\mathcal{Z}^{2,2}_{k=3}&=&(|A|^2+|D|^2)+2(AB^*+DC^*), 
\end{eqnarray}
where
\begin{eqnarray}
A\equiv\chi_0+\chi_3;\,\,B\equiv\chi_{2/3};\,\,C\equiv\chi_{1/15};\,\,D\equiv\chi_{2/5}+\chi_{7/5}.
\end{eqnarray}

\subsection{Parafermionic minimal model dual to a tricritical three-state Potts model}
We apply the parafermionization~(\ref{parafermionization_1}) to the partition functions~(\ref{tripotts_0},\ref{tripotts_1},\ref{tripotts_2}) of a tricritical three-state Potts model to obtain the last parafermionic minimal model: 
\begin{eqnarray}
{\mathcal{Z}'_3}^{0,0}={\mathcal{Z}'_3}^{1,1}&=&\omega^*(A'{D'}^*+B'{E'}^*+C'{F'}^*-|D'|^2-|E'|^2-|F'|^2);\\
{\mathcal{Z}'_3}^{0,1}={\mathcal{Z}'_3}^{1,0}&=&\omega(A'{D'}^*+B'{E'}^*+C'{F'}^*-|D'|^2-|E'|^2-|F'|^2);\\
{\mathcal{Z}'_3}^{0,2}={\mathcal{Z}'_3}^{1,2}&=&A'{D'}^*+B'{E'}^*+C'{F'}^*+2(|D|^2+|E|^2+|F|^2); \\
{\mathcal{Z}'_3}^{2,0}={\mathcal{Z}'_3}^{2,1}&=&(|A'|^2+|B'|^2+|C'|^2)-\left(D'{A'}^*+E'{B'}^*+F'{C'}^*\right);\\
{\mathcal{Z}'_3}^{2,2}&=&(|A'|^2+|B'|^2+|C'|^2)+2\left(D'{A'}^*+E'{B'}^*+F'{C'}^*\right), 
\end{eqnarray}
where
\begin{eqnarray}
A'\equiv\chi_0+\chi_5;\,B'\equiv\chi_{1/7}+\chi_{22/7};\,C'\equiv\chi_{5/7}+\chi_{12/7};\,D'\equiv\chi_{4/3};\,E'\equiv\chi_{10/21};\,F'\equiv\chi_{1/21}.\nonumber
\end{eqnarray}

\sloppy

\providecommand{\href}[2]{#2}\begingroup\raggedright\endgroup

\end{document}